\documentclass[twocolumn,preprintnumbers,amsmath,amssymb]{revtex4}

\usepackage{graphicx}  
\usepackage{sidecap}
\begin{document}
\title{Heat capacity of nonequilibrium electron-hole plasma in graphene layers and graphene~bilayers}
\author{V. Ryzhii$^{1,2,3}$,  M. Ryzhii$^4$,  T.~Otsuji$^1$,   
V. Mitin$^5$, and M. S. Shur$^6$}
\address{
$^1$Research Institute of Electrical Communication, Tohoku University, Sendai 980-8577, Japan\\
$^2$Institute of Ultra High Frequency Semiconductor Electronics of RAS,\\
 Moscow 117105, Russia\\
 $^3$Center for Photonics and Two-Dimensional Materials, Moscow Institute of Physics
and Technology, Dolgoprudny 141700, Russia\\
 $^4$Department of Computer Science and Engineering, University of Aizu, Aizu-Wakamatsu 965-8580, Japan\\
 $^5$Department of Electrical Engineering,~University at Buffalo, SUNY, Buffalo, New York 1460-1920 USA\\
$^6$Department of Electrical, Computer, and Systems Engineering, Rensselaer Polytechnic Institute, Troy, New York 12180, USA\\
}
 \begin{abstract} 
 We analyze the statistical characteristics of
   the quasi-nonequilibrium two-dimensional electron-hole plasma in
 graphene layers (GLs) and graphene bilayers (GBLs) and evaluate their heat capacity. The GL heat capacity of the weakly pumped intrinsic or  weakly doped  GLs normalized by the Boltzmann constant is  equal to 
$c_{GL} \simeq 6.58$. With varying carrier temperature  the intrinsic GBL carrier heat capacity  $c_{GBL}$ changes  from $c_{GBL} \simeq 2.37$  at $T \lesssim 300$~K to
$c_{GBL} \simeq 6.58$ at elevated temperatures. These values are markedly different
from the  heat capacity of classical two-dimensional carriers with $c = 1$.
 The obtained results can be useful for the optimization of different GL- and GBL-based high-speed devices. 
\end{abstract}

\maketitle

\newpage

\section{Introduction}

The properties of the graphene layers (GLs) and graphene bilayers (GBLs),
in particular, their optical characteristics, conductivity, plasmonic properties,  thermal conductivity (both associated with the lattice and the carriers), heat capacity, and others  have been extensively studied theoretically and experimentally~\cite{1,2,3,4,5,6,7,8,9,10,11,12,13,14,15,16,17,18,19} (see the references therein). The contributions of the carriers in GLs and GBLs to the overall heat capacity is small in comparison with
the contribution of the lattice vibrations~\cite{20}. However,  the electron and hole  heat capacity determines the rate of the carrier  heating and cooling. This 
heating/cooling rate affects for
the ultimate high-speed performance, including  the dynamic response and the modulation  characteristics of the GL- and GBL-based devices using
the variation of the two-dimensional electron-hole plasma (2DEHP) parameters
(such as the effective carrier temperature, conductivity, transparency of the incident
radiation). Such GL- and GBL-devices include the
 carrier bolometric detectors~\cite{21,22}, electro-optical modulators~\cite{23}, fast thermal radiation emitters~\cite{24,25,26,27,28,29,30,31}, and  superluminescent and lasing diodes~\cite{32}.  
Many papers deal with the theoretical evaluation of the GL- and GBL-lattice heat capacity (see, for example, a recent review~\cite{33}). However, 
the carrier capacity of the intrinsic quasi-nonequilibrium 2DEHP in  GLs and GBLs was not addressed. The case of highly doped GLs was briefly discussed in ~\cite{34,35}.
 In this paper, we  calculate the heat capacity of the 2DEHP in the equilibrium 
 and of the 2DEHP somewhat deviating from the equilibrium due to
the radiation absorption and/or the carrier injection.

\section{General relations}
 The dispersion relations for electrons (upper sign) and holes (lower sign)
 in the GLs and GBLs are presented as
 
\begin{eqnarray}\label{eq1}
\varepsilon_{GL}^{\pm} = \pm v_Wp, \qquad
\varepsilon_{GBL}^{\pm} \simeq \pm\frac{\gamma_1}{2}
\biggl[\sqrt{1 + 4v_W^2p^2/\gamma_1^2}- 1\biggr].
\end{eqnarray}
Here $\hbar$ is the Planck constant, $v_W \simeq 10^8$~cm/s is the characteristic carrier velocity in GL and GBLs, $p = |p|$ is the carrier momentum, and $\gamma_1 \simeq 0.4$~eV is the band parameter (the GBL hopping parameter)~\cite{36,37}.

We assume that the frequent carrier-carrier collisions lead to the formation
in the 2DEHP of the electron and hole systems described by quasi-Fermi energy distribution
functions $f_e(\varepsilon)$ and $f_h(\varepsilon)$ with common effective temperature $T$ 
:
$f_e(\varepsilon) = \displaystyle \biggl[1 + \exp\biggl(\frac{\varepsilon - \mu_e}{k_BT}\biggr) \biggr]^{-1}$ and $f_h(\varepsilon)= \displaystyle \biggl[1 + \exp\biggl(\frac{\varepsilon - \mu_h}{k_BT}\biggr) \biggr]^{-1}$, where $k_B$ is the Boltzmann constant, $\varepsilon \geq 0$ is the carrier kinetic energy and $\mu_e$ and $\mu_h$
are the electron and hole quasi-Fermi energies, respectively. In the undoped GLs and GBLs, $\mu_e = \mu_h =0$. If, in particular,  the GL (or the GBL) is doped by donors, $\mu_e > 0$, while $\mu_h <0$. In the equilibrium, i.e., without optical or injection pumping and with no heating of the 2DEHP by the electric field, $\mu_e + \mu_h = 0$.
In this case, $\mu_e$ and, consequently, $\mu_h$ are determined by the donor sheet
density $\Sigma_d$. In the acceptor doped GLs (GBLs), $\mu_h > 0$ and $\mu_e <0$
with $\mu_h$ and $\mu_e$ determined by the acceptor density $\Sigma_a$.
When the 2DEHP is off equilibrium, generally, $\mu_e + \mu_h \neq 0$. 

When the 2DEHP deviates from equilibrium due the irradiation or the carrier injection, the combined quasi-Fermi energy $\mu = \mu_e + \mu_h$ can be either positive or negative (see, for example,~\cite{14}). The main mechanism, which enables tending of $\mu$ zero, is the Auger recombination. However, in the 2DEHP under consideration
the Auger recombination-generation processes are relatively ineffective~\cite{14,38,39}.

The net carrier (electrons and holes) densities, $\Sigma_{GL}$ and $\Sigma_{GBL}$,  in the GL and GBL, respectively, in line with Eq.~(1) are given by

\begin{eqnarray}\label{eq2}
 \Sigma_{GL} = \frac{2}{\pi\hbar^2v_W^2}\int_0^{\infty}d\varepsilon \varepsilon\nonumber\\
 \times\biggl[\frac{1}{1 + \displaystyle\exp\biggl(\frac{\varepsilon - \mu_e}{k_BT}\biggr)}
 + \frac{1}{1 + \displaystyle\exp\biggl(\frac{\varepsilon - \mu_h}{k_BT}\biggr)}\biggr]
 \nonumber\\
= \frac{2(k_BT)^2}{\pi\hbar^2v_W^2}\biggl[{\cal F}_1\biggl(\frac{\mu_e}{k_BT}
 \biggr) + {\cal F}_1\biggl(\frac{\mu_h}{k_BT}\biggr)\biggr]
\end{eqnarray}
 and

\begin{eqnarray}\label{eq3}
\Sigma_{GBL}  = \frac{2}{\pi\hbar^2v_W^2}\int_0^{\infty}d\varepsilon (\varepsilon + \gamma_1/2)\nonumber\\
\times\biggl[\frac{1}{1 + \displaystyle\exp\biggl(\frac{\varepsilon - \mu_e}{k_BT}\biggr)}
+\biggl[\frac{1}{1 + \displaystyle\exp\biggl(\frac{\varepsilon - \mu_h}{k_BT}\biggr)}\biggr]
\nonumber\\
=\frac{2}{\pi\hbar^2v_W^2}\biggl[{\cal F}_1\biggl(\frac{\mu_e}{k_BT}\biggr)
 + {\cal F}_1\biggl(\frac{\mu_h}{k_BT}\biggr)\nonumber\\
 +\frac{\gamma_1}{k_BT}{\cal F}_0\biggl(\frac{\mu_e}{k_BT} +
 \biggr) + \frac{\gamma_1}{k_BT}{\cal F}_0\biggl(\frac{\mu_h}{k_BT}\biggr)\biggr].
\end{eqnarray}
Here ${\cal F}_{\xi}(y)$ is the Fermi-Dirac integral.

 The  carrier energy in the 2DEHP can be calculated as

\begin{eqnarray}\label{eq4}
{\cal E}_{GL} =\frac{2}{\pi\hbar^2v_W^2}\int_0^{\infty}d\varepsilon\varepsilon^2\nonumber\\
\times\biggl[\frac{1}{1 + \displaystyle\exp\biggl(\frac{\varepsilon - \mu_e}{k_BT}\biggr)}
 + \frac{1}{1 + \displaystyle\exp\biggl(\frac{\varepsilon - \mu_h}{k_BT}\biggr)}\biggr]\nonumber\\
 = \frac{2(k_BT)^2}{\pi\hbar^2v_W^2}\biggl[{\cal F}_2\biggl(\frac{\mu_e}{k_BT}
 \biggr) + {\cal F}_2\biggl(\frac{\mu_h}{k_BT}\biggr)\biggr]
\end{eqnarray}
and

\begin{eqnarray}\label{eq5}
{\cal E}_{GBL} =\frac{2}{\pi\hbar^2v_W^2}\int_0^{\infty}d\varepsilon\varepsilon(\varepsilon + \gamma_1/2)\nonumber\\
\times\biggl[\frac{1}{1 + \displaystyle\exp\biggl(\frac{\varepsilon - \mu_e}{k_BT}\biggr)}
 + \frac{1}{1 + \displaystyle\exp\biggl(\frac{\varepsilon - \mu_h}{k_BT}\biggr)}\biggr]\nonumber\\
=\frac{2}{\pi\hbar^2v_W^2}\biggl[{\cal F}_2\biggl(\frac{\mu_e}{k_BT}\biggr)
 + {\cal F}_2\biggl(\frac{\mu_h}{k_BT}\biggr)\nonumber\\
 +\frac{\gamma_1}{k_BT}{\cal F}_1\biggl(\frac{\mu_e}{k_BT}\biggr) + \frac{\gamma_1}{k_BT}{\cal F}_1\biggl(\frac{\mu_h}{k_BT}\biggr)\biggr]. 
\end{eqnarray}

\section{Quasi-nonequilibrium 2DEHP in  weakly doped GL and GBL}

For a weakly nonequilibtium 2DEHP  
Eqs.~(2) and (3) yield
well known formulas for the carrier densities in GLs and GBLs:

\begin{eqnarray}\label{eq6}
 \Sigma_{GL} = \biggl(\frac{k_BT}{\hbar\,v_W}\biggr)^2\biggl(\frac{\pi}{3} + \frac{4\ln 2}{\pi} \frac{\mu}{k_BT}\biggr),
\end{eqnarray}

\begin{eqnarray}\label{eq7}
 \Sigma_{GBL} = \biggl(\frac{k_BT}{\hbar\,v_W}\biggr)^2\biggl[\frac{\pi}{3} 
+ \frac{2\ln 2}{\pi}\frac{\gamma_1}{k_BT}
\nonumber\\
+\biggl(\frac{4\ln2}{\pi}  +  \frac{\gamma_1}{k_BT}\biggr) \frac{\mu}{k_BT}\biggr].
\end{eqnarray}
Equations~(4) and (5) result in the following expressions for
the carrier thermal energy density (thermal energy per GL and GBL area) as a function of the carrier effective temperature $T$
and the combined quasi-Fermi energy $\mu$:

\begin{eqnarray}\label{eq8}
{\cal E}_{GL} \simeq \frac{2(k_BT)^3}{\pi\hbar^2v_W^2} \biggl[3\zeta(3) + \frac{\pi^2}{3}\frac{\mu}{k_BT}\biggr],
\end{eqnarray}

\begin{eqnarray}\label{eq9}
{\cal E}_{GBL} \simeq \frac{2(k_BT)^3}{\pi\hbar^2v_W^2} \biggl[3\zeta(3) + \frac{\pi^2}{12}\frac{\gamma_1}{k_BT}
\nonumber\\
 + \biggl(\frac{\pi^2}{3} + \ln 2\frac{\gamma_1}{k_BT}\biggr)\frac{\mu}{k_BT}\biggr],
\end{eqnarray}
where  $\zeta(x)$ is the Riemann zeta function: $\zeta(3) \simeq 1.202$.

Considering that the 2DEHP heat capacities in GLs and  GBLs (per area) are defined as
$C_{GL} = d {\cal E}_{GL}/d T$ and $C_{GBL} = 
d {\cal E}_{GBL}/d T$, we obtain from Eqs.~(8) and (9)

\begin{eqnarray}\label{eq10}
C_{GL} \simeq \frac{2(k_BT)^2}{\pi\hbar^2v_W^2} \biggl[9\zeta(3) + \frac{2\pi^2}{3}\frac{\mu}{k_BT}\biggr]\nonumber\\
\simeq \frac{6.58\pi}{3}\biggl(\frac{k_BT}{\hbar\,v_W}\biggr)^2, 
\end{eqnarray}

\begin{eqnarray}\label{eq11}
C_{GBL} \simeq \frac{2(k_BT)^2}{\pi\hbar^2v_W^2} \biggl[9\zeta(3) + 
\frac{\pi^2}{6}\frac{\gamma_1}{k_BT}
\nonumber\\
 + \biggl(\frac{2\pi^2}{3} + \ln 2\frac{\gamma_1}{T}\biggr)\frac{\mu}{k_BT}\biggr]
 \nonumber\\
\simeq \frac{\pi}{3}\biggl(\frac{k_BT}{\hbar\,v_W}\biggr)^2
\biggl(6.57 + \frac{\gamma_1}{k_BT}\biggr).
\end{eqnarray}

Since at $\mu =0$, according to Eqs.~(6) and (7),

\begin{eqnarray}\label{eq12}
 \Sigma_{GL} \simeq \frac{\pi}{3} \biggl(\frac{k_BT}{\hbar\,v_W}\biggr)^2,
\end{eqnarray}

\begin{eqnarray}\label{eq13}
 \Sigma_{GBL} \simeq \frac{\pi}{3}\biggl(\frac{k_BT}{\hbar\,v_W}\biggr)^2\biggl(1 + \frac{6\ln 2}{\pi^2}\frac{\gamma_1}{k_BT}\biggl),
\end{eqnarray}
the pertinent heat capacitances, $c_{GL}= C_{GL}/k_B\Sigma_{GL}$ and $c_{GBL}= C_{GBL}/k_B\Sigma_{GBL}$ (normalized by $k_B$, i.e., in units of the Boltzman constant), per one carrier are equal to

\begin{eqnarray}\label{eq14}
 c_{GL} \simeq \frac{54\zeta(3)}{\pi^2} \simeq 6.58,
\end{eqnarray}
\begin{eqnarray}\label{eq15}
  c_{GBL} \simeq \frac{\pi^2}{6\ln\,2}\Biggl(\frac{1 + \displaystyle \frac{54\zeta(3)}{\pi^2}\frac{k_BT}{\gamma_1}}{1 +\displaystyle\frac{\pi^2}{6\ln\,2}\frac{k_T}{\gamma_1}}\Biggr)
\nonumber\\
   \simeq 2.37\Biggl(\frac{1 + \displaystyle 6.58\frac{k_BT}{\gamma_1}}{1 +\displaystyle 2.37\frac{k_BT}{\gamma_1}}\Biggr).
\end{eqnarray}

\section{Comments}

\begin{figure}[t]
\centering
\includegraphics[width=7.0cm]{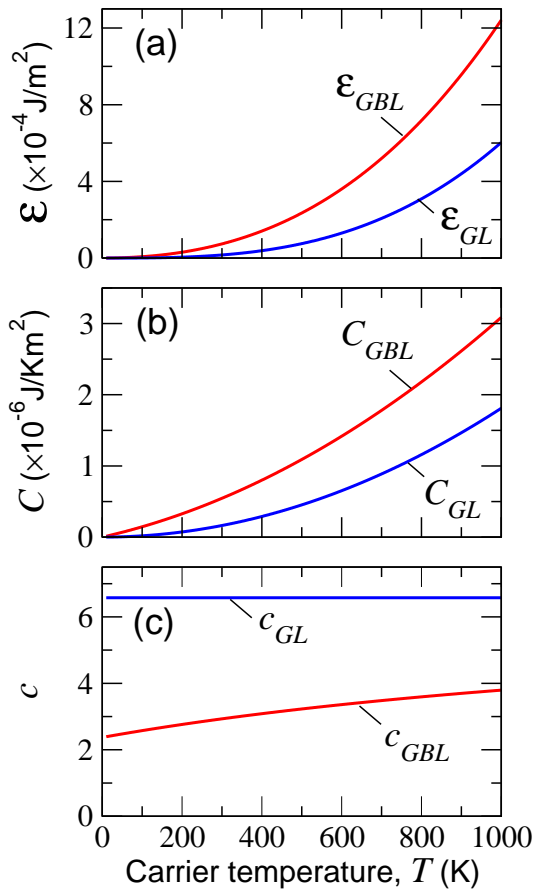}
\caption{ The  carrier (a) thermal  energies  ${\cal E}_{GL}$ and ${\cal E}_{GL}$ per area, (b) heat capacities $C_{GL}$ and $C_{GBL}$ per area, and (c) heat capacities per one carrier $c_{GL}$ and $c_{GBL}$, normalized by $k_B$, versus carrier temperature $T$ at $\mu = 0$ ($\gamma_1 = 0.4$~eV).}
\label{F1
}
\end{figure}

At $k_BT \ll \gamma_1$ ($T \lesssim 300$~K), Eq.~(13) yields $c_{GBL} \simeq (\pi^2/6\ln 2) \simeq 2.37 $.
When $T$ is rather high, $c_{GBL}$ increases tending to $c_{GBL} \simeq 6.58$.
Figure~2 shows the temperature dependences of the energy densities ${\cal E}_{GL}$ and ${\cal E}_{GL}$ and the  heat capacities per one carrier $c_{GL}$ and $c_{GBL}$
calculated using Eqs.~(10), (11), (14), and (15) for $\mu = 0$ (equilibrium 2DEHP) assuming $\gamma_1 = 0.4$~eV.

A  noticeable deviation of $c_{GL}$ and $c_{GBL}$ from the classical value for nondegenerate 2D systems (ie., from $c = 1$) seen from Eqs.~(14) and (15) and from Fig.~2(c), is associated with the nonparabolicity of the carrier spectra in both GLs and GBLs. The nonparabolicity    provides different densities of states (a linear in GLs and a linear rising from a constant at the Dirac point in GBLs), whereas the absence of the energy gap leads to
a weak degeneration near the Dirac point ($f_e(0) = f_h(0) \simeq 1/2$). 
In particular, if we would neglect the partial degeneracy  effect, we obtain $c_{GL} = 6$
and $ 1 <c_{GBL} = (1 + 6k_BT/\gamma_1)/(1 + k_BT/\gamma_1) < 6$, respectively.

The variation of $\mu$ with the effective carrier temperature leads to a small modification
of $c_{GL}$ and $c_{GBL}$ assuming a weak deviation from equilibrium. Depending on the pumping or heating conditions, this effect can result in either  increase or a decrease in  
$\mu$ (see, for example,~\cite{14}) and, hence, in somewhat  varying of $c_{GL}$ and $c_{GBL}$.

In the case when the gapless carrier density of state is given by a power energy dependence
$\rho(\varepsilon) \propto \varepsilon^{\xi}$,
for the heat capacity per a carrier $c_{\xi}$ one can obtain

\begin{eqnarray}\label{eq16}
c_{\xi} = (\xi +2)\frac{\int_0^{\infty}dx x^{\xi + 1}[1 + \exp(x-\mu/k_BT)]^{-1}}{\int_0^{\infty}dx x^{\xi}[1 + \exp(x - \mu/k_BT)]^{-1}}\nonumber\\
 = (\xi +2)\frac{{\cal F}_{\xi+1}(\mu/k_BT)}{{\cal F}_{\xi}(\mu/k_BT)}.
\end{eqnarray}
In particular, at $\mu = 0$, Eq.~(16) yields

\begin{eqnarray}\label{eq17}
c_{\xi} = 
 (\xi +2)\frac{{\cal F}_{\xi+1}(0)}{{\cal F}_{\xi}(0)}\nonumber\\
  =  (\xi +2)\frac{\Gamma(\xi+2)}{\Gamma(\xi+1)}\frac{\zeta(\xi+2)}{\zeta(\xi+1)}\biggl[\frac{1 - 1/2^{(\xi+1)}}
 {1 - 1/2^{\xi}}\biggr],
\end{eqnarray}
where $\Gamma(x)$ is the Gamma function.
For GLs ($\xi = 1$) and GBLs ($\xi = 0$, $k_BT \ll \gamma_1$),
from Eq.~(17) we obtain $c_{GL} = c_1 = 54\zeta(3)/\pi^2$ and
$c_{GBL} \simeq  c_0  = \pi^2/6\ln 2$, that 
 actually coincides with Eqs.~(14) and (15).

The renormalization of the carrier spectrum and the density of state energy dependence in GLs, associated with the carrier-carrier interactions 
(for example,~\cite{6,40,41,42,43}), affects the GL heat capacity. 
To estimate the role  the Fermi-liquid effect in GLs associated with the inter-carrier
interaction,  following~\cite{43}, in comparison with Eq.~(1) we modify the carrier dispersion law in GLs as follows:

\begin{eqnarray}\label{eq18}
 \varepsilon^{\pm}_{GL} = \pm v_Wp\biggl[1 + g\ln\biggl(\frac{{\cal K}\hbar}{p}\biggr)\biggr].
\end{eqnarray}
Here $g = e^2/(8\pi\hbar\,v_W\kappa)$ is the dimensionless carrier-carrier interaction parameter, where $\kappa$ is the effective dielectric constant, and ${\cal K}$ is the cut-off parameter~\cite{6,40,41} (${\cal K} \simeq 0.5\times 10^8$~cm$^{-1}$).

Considering Eq.~(16), i.e., accounting for the carrier velocity renormalization,  
at $\mu = 0$ for the renormalized carrier density $\Sigma_{GL}^*$, density of the carrier energy ${\cal E}_{GL}^*$, and the carrier heat capacity per one carrier $c_{GL}$ we obtain

\begin{eqnarray}\label{eq19}
r = \frac{\Sigma_{GL}^*}{\Sigma_{GL}} \simeq \frac{{\cal E}_{GL}^*}{{\cal E}_{GL}}
\simeq  \biggl[1 + g\ln\biggl(\frac{{\cal K}\hbar\,v_W}{k_BT}\biggr)\biggr]^{-2} < 1
\end{eqnarray}
and 
\vspace{-10mm}

\begin{eqnarray}\label{eq20}
c_{GL}^* \simeq c_{GL}. 
\end{eqnarray}
Setting $\kappa = 2.5$, at $T = (10 - 300)$~K, we obtain $r \simeq 0.59 - 0.72$.
One can see from Eq.~(19) that the inclusion of the the Fermi-liquid effect 
results in natural lowering of the thermal carrier energy (due to a decrease in the density of states near the Dirack point), but, according to Eq.~(20), this
does not lead to a change in $c_{GL}$.\\

\section{Conclusions}

\vspace{-3mm}
We calculated the heat capacity per one carrier of the quasi-equilibrium 2DEHP in GLs and GBLs and demonstrated that it  can be larger from its classical values.
The speed of operation (the switching time, turn-on time, and maximum modulation frequency) of the GL- and GBL-based
devices, such as the bolometric photodetectors of
the terahertz and infrared radiation, 
electro-optical modulators, fast thermal radiation emitters, and superluminescent 
and lasing diodes is affected by the carrier heating/cooling and is determined by the 
product 
 of $c_{GL}$
or $c_{GBL}$ and the carrier energy relaxation time. Therefore, our results are important for  the evaluation of the ultimate characteristics and   optimization of such  devices.

\section*{Acknowledgments}

The work
at RIEC and UoA was supported by the Japan Society for Promotion of Science (KAKENHI Grant No.~16H02336) and the RIEC Nation-Wide Collaborative Research Project No.~H31/A01, Japan.
The work at RPI was supported by the Office of Naval Research (Project Manager Dr. Paul Maki), the US Air Force Office of Scientific Research (FA9550-19-1-0355, Project Manager
Dr. John Qiu), and the Army Research Laboratory under ARL MSME Alliance (Project Manager D. Meredith Reed), USA.

\end{document}